\documentclass[intlimits,twoside,a4paper]{article}

\usepackage{amsmath,amssymb} 
\usepackage{graphicx}        

\usepackage{bm}              

\usepackage[T2A,T1]{fontenc} 
\usepackage[cp1251]{inputenc}

\usepackage[eqsecnum]{cmpj3}

\issue{2021}{24}{1}{13301}
\doinumber{10.5488/CMP.24.13301}
\title[Frictional anisotropy of palladium nanoparticles on graphene]%
{Atomistic modelling of frictional anisotropy of palladium nanoparticles on graphene%
}
\author[A.V. Khomenko, M.V. Zakharov]{A.V. Khomenko\refaddr{label1,label2},
        M.V. Zakharov\refaddr{label1}}
\addresses{
\addr{label1} Sumy State University, 2 Rimsky-Korsakov St., 40007 Sumy, Ukraine
\addr{label2} Peter Gr\"unberg Institut-1, Forschungszentrum-J\"ulich, D-52425 J\"ulich, Germany
}
\date{Received June 30, 2020, in final form September 10, 2020}

\begin{document}

\maketitle

\begin{abstract}
This article is a continuation of our previous studies of the frictional anisotropy of metal nanoparticles on the surface of a graphene substrate for other temperature conditions. The friction force acting on palladium nanoparticles on a graphene sheet in various lateral directions is investigated using classical molecular dynamics modelling. Anisotropy is studied at high sliding speeds of nanoparticles consisting of 10000 atoms on the surface of graphene. The effect of incommensurability and short-range order of the contact surfaces of nanoparticles, as well as the graphene deformation lead to the absence of an expressed angular dependence of the friction force.
\end{abstract}

\section{Introduction}

The friction force is responsible for a variety of phenomena in our everyday life including playing the violin, dancing, car driving, etc. The same wide range of using the friction can be found in nanobiotechnology~\cite{b1,Pogrebnjak_2012}. The anisotropy of friction is the dependence of rubbing on the relative orientation of two crystalline surfaces creating a solid interface. This difficultly is encountered in the majority of engineering applications, and it is the fundamental tribological phenomenon. New technologies such as micromechanical systems require an accurate friction control at the molecular level, because the lubricating films and coatings used in these technologies, have the width of several molecular layers. The friction between surfaces in sliding contact is dependent on their structure and orientation relatively to each other and to the direction of sliding. The frictional anisotropy can be explored by changing the shear direction of two surfaces with a fixed relative orientation or by fixing the shear direction and changing the relative crystallographic orientation of two surfaces.
Two surfaces that contact each other can be exposed either to elastic or plastic deformation \cite{b2}. In general, the anisotropy is of two types: inherent and induced. The first is usually the result of sedimentation of particles, while the second is formed during the plastic deformation process, for example, in the process of a structural change of a material or forming of oriented microcracks \cite{b3}.
Usually, metal surfaces undergo a plastic deformation under contact conditions. The research of frictional anisotropy between plastically deformed surfaces of single crystalline metals suggests that the commensurability effects of surface lattices are not its main reason. There is an assumption that the anisotropy of friction is mainly  caused by slipping of the atomic planes in the volume of a crystal and it is not conditioned by commensurability of the surface lattice \cite{b2}. The frictional anisotropy is also present in non-metallic solids, such as a diamond or a sapphire. Such materials exhibit high track widths and high friction coefficients, whereas for metals there are narrow tracks.
Two problems related with a trace width can be found in the literature on anisotropy. First, the track width is not measured accurately; second, the plowing term of adhesion theory is not perceived as the cause of anisotropy although the sizes of the tips used were always very small. If the frictional heating is not  taken into account, the rubbing resistance is considered to be formed of two parts: the interfacial shear resistance between the friction surface and the slipping block and the resistance resulting from the sliding plowing. In this regard, the mechanical characteristics such as hardness, yield and tensile stresses are very important for any friction mechanism interpretation between solids \cite{b4,Troshchenko_CMP2015}.
With the help of small loads of the order of $10^{-8}$ N, the strain of the solid surfaces is minimized. However, if the load is reduced so that the elastic strain  ceases to be observed, then the friction anisotropy will disappear. In other words, the anisotropy of friction can be observed between the surfaces that are elastically deformed while sliding \cite{b2}.
When the contact pressure exceeds the critical value, the frictional anisotropy arises. This phenomenon is explained by surface and subsurface failure that is formed in the preferred crystallographic directions during sliding \cite{b5,Pogr_Gonch_MFiNT_16}. The formation of some plastic groove is possible in the friction track, although the pressure of 20 GPa seems to be too low for this. On the other hand, some microplastic deformation can boost the energy dissipation, but the main factor is a subsurface cracking that is formed when the contact pressure exceeds some critical value \cite{b6}.

In our previous works \cite{TL_19_Khom,jnep_13}, based on the classical molecular dynamics, we describe the anisotropy of friction of aluminum, platinum, palladium, silver, and nickel nanoparticles adsorbed on a graphene layer. The present study is devoted to the analysis of this phenomenon for palladium nanoparticles and for different temperature variations.

\section{Methodology}

\begin{figure}[!b]
\centerline{\includegraphics[width=0.45\textwidth]{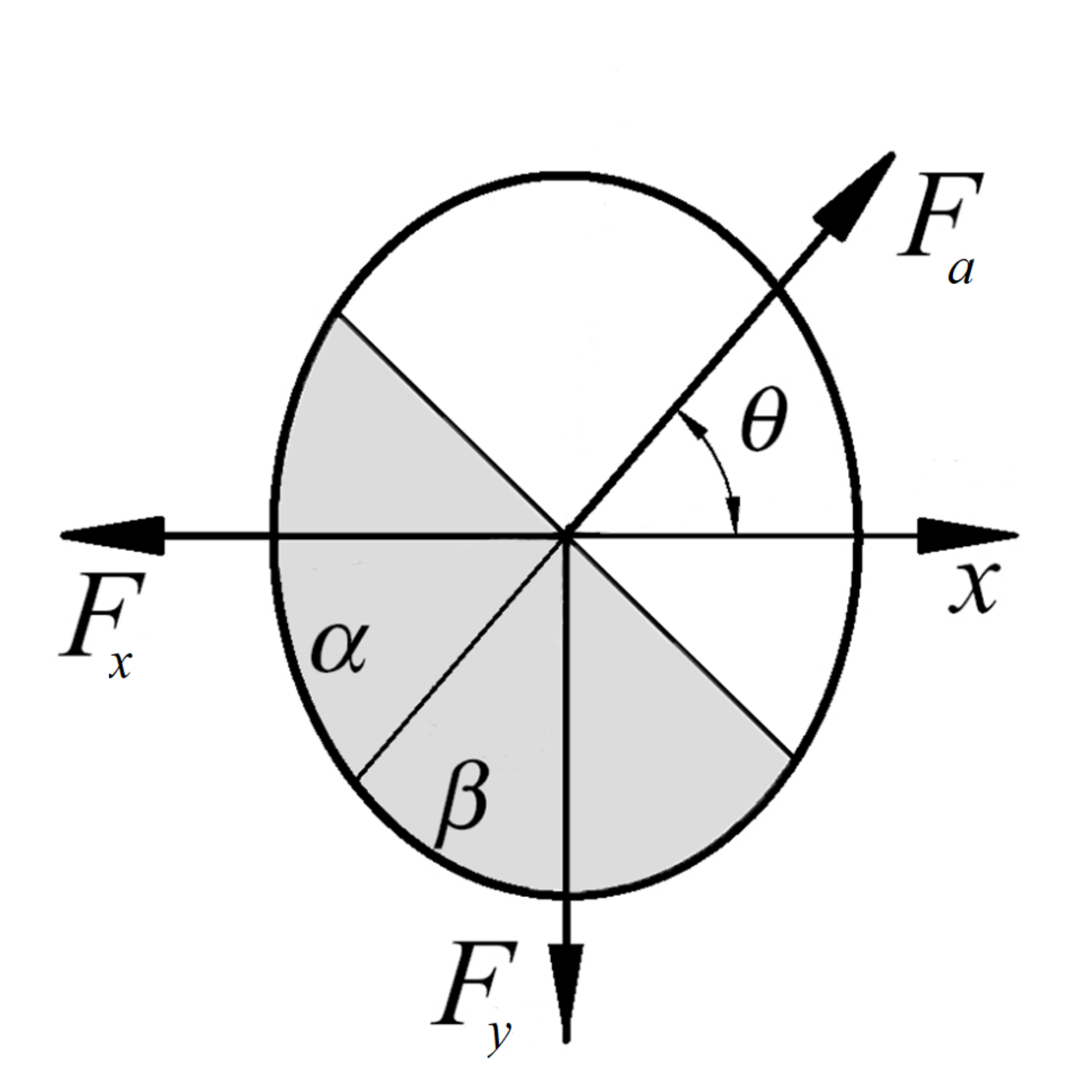}\includegraphics[width=0.45\textwidth]{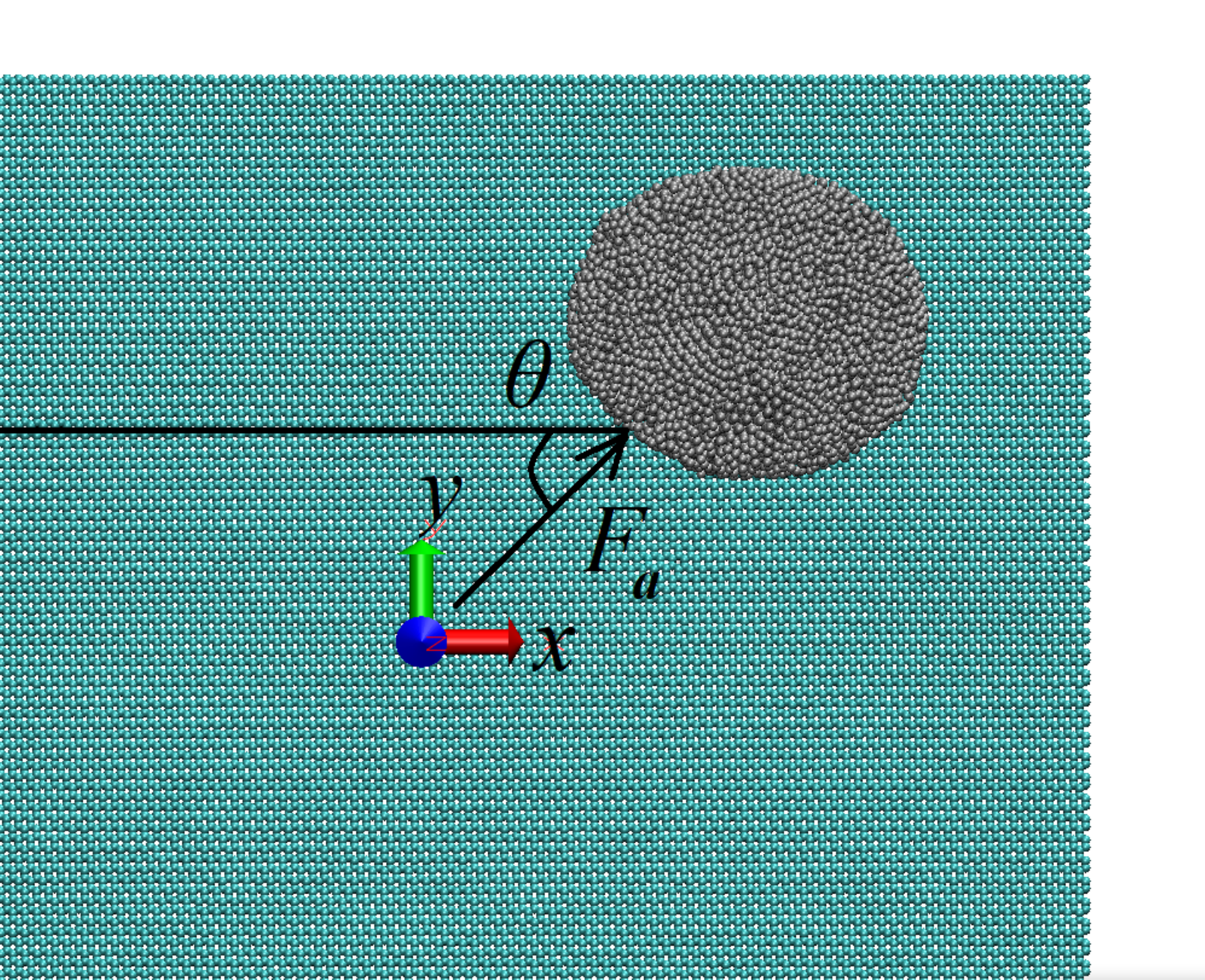}}
\caption{(Colour online) Driving forces acting on the nanoparticle during sliding on graphene substrate (left-hand) and top view of the system with sliding nanoparticle at $\theta = 45^\circ$ (right-hand).} \label{fig1}
\end{figure}

The calculations were carried out using the program created in the Microsoft Visual Studio 2008 development environment using parallel computing technology Nvidia CUDA. In our calculations, we simulate the following system: a nanoparticle consisting of 10000 palladium atoms and moving along the surface of graphene, which consists of $64\times 64$ cells and 65536 carbon atoms. The graphene substrate has fixed zigzag and armchair edges along the $x$ and $y$ axes and is parallel to the $x$ and $y$ plane (figure~\ref{fig1}). During the simulation the substrate remains stationary only along the edges.

Embedded atom model potential is used to describe the interaction of palladium atoms \cite{b7}. The carbon interaction forces in a graphene substrate are represented by a harmonic potential \cite{b8}. The Lennard–Jones potential with the parameters of the energy $\epsilon = 0.008738$~eV and length  $\sigma = 0.24945$~nm depicts the interatomic energy between palladium atoms of a nanoparticle and carbon atoms of a graphene substrate \cite{Khomenko2010,b10,Khomenko2013}. Calculations are carried out in 3 stages. During the first stage,  there occurs a destruction  of the face-centered cubic lattice of  one layer of palladium atoms as a result of heating. The metal becomes amorphous. The next step is the formation of a nanoparticle. Nanoparticle is cooled through the use of a Berendsen thermostat due to which the excess heat is removed and the preset temperature is reached~\cite{b11}. At the same time, the formation of a polycrystalline nanoparticle structure occurs~\cite{Yushchenko_UJP2011,Stef_PTT_19,BJ2018_rev}. The last step is the movement of a nanoparticle through the use of force $F_a$.

The shear force is applied to each atom located in the selected sector in figure~\ref{fig1}, in the direction determined by the angle of $\theta$ counted from the $x$-direction. Sectors are formed and divided by the diagonal line $F_a$ and the angles $\alpha$ = $\beta$ = $90^\circ$ hold approximately the same number of atoms in each half of the sector, which is symmetrical to the direction of the applied force $F_a$. Zero torque relatively to the center of mass of the nanoparticle in the $x$ - $y$ plane is provided by the geometry described above, which prevents the rotation of the nanoparticle in this plane. We do not simulate the rotation of a nanoparticle in order to provide a rectilinear motion along the surface of graphene. The  resultant frictional force acting on a nanoparticle is determined as follows:
\begin{align}
F=\sqrt{F_{x}^2+F_{y}^2}\,,
\end{align}
where $F_x$ and $F_y$ are the friction force components along the corresponding axes. Frictional force is measured for the following angles: $15^\circ$, $22.5^\circ$, $30^\circ$, $37.5^\circ$, $45^\circ$ and $60^\circ$.

\section{Results}

\begin{figure}[!b]
\centerline{\includegraphics[width=0.45\textwidth]{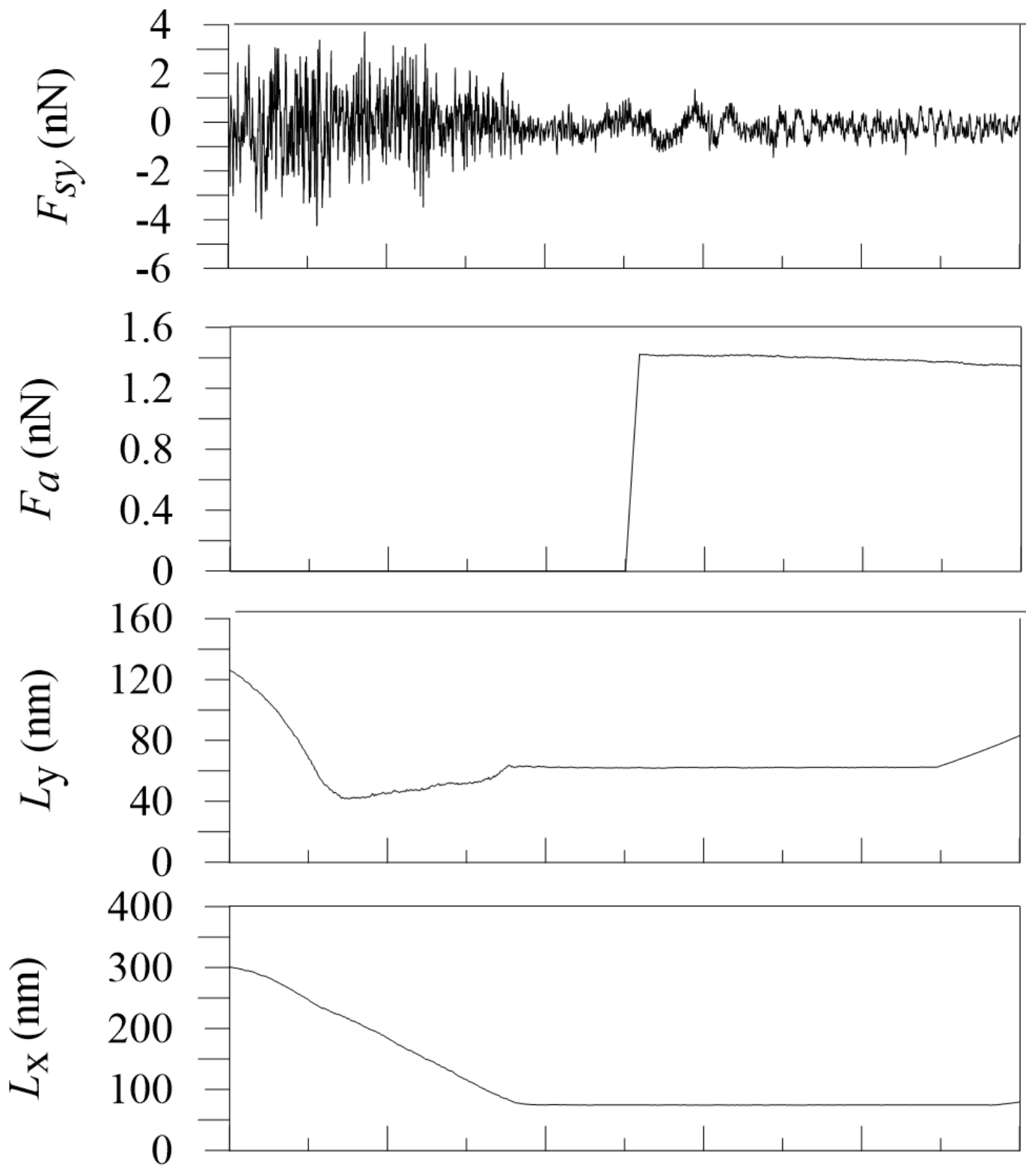}\includegraphics[width=0.45\textwidth]{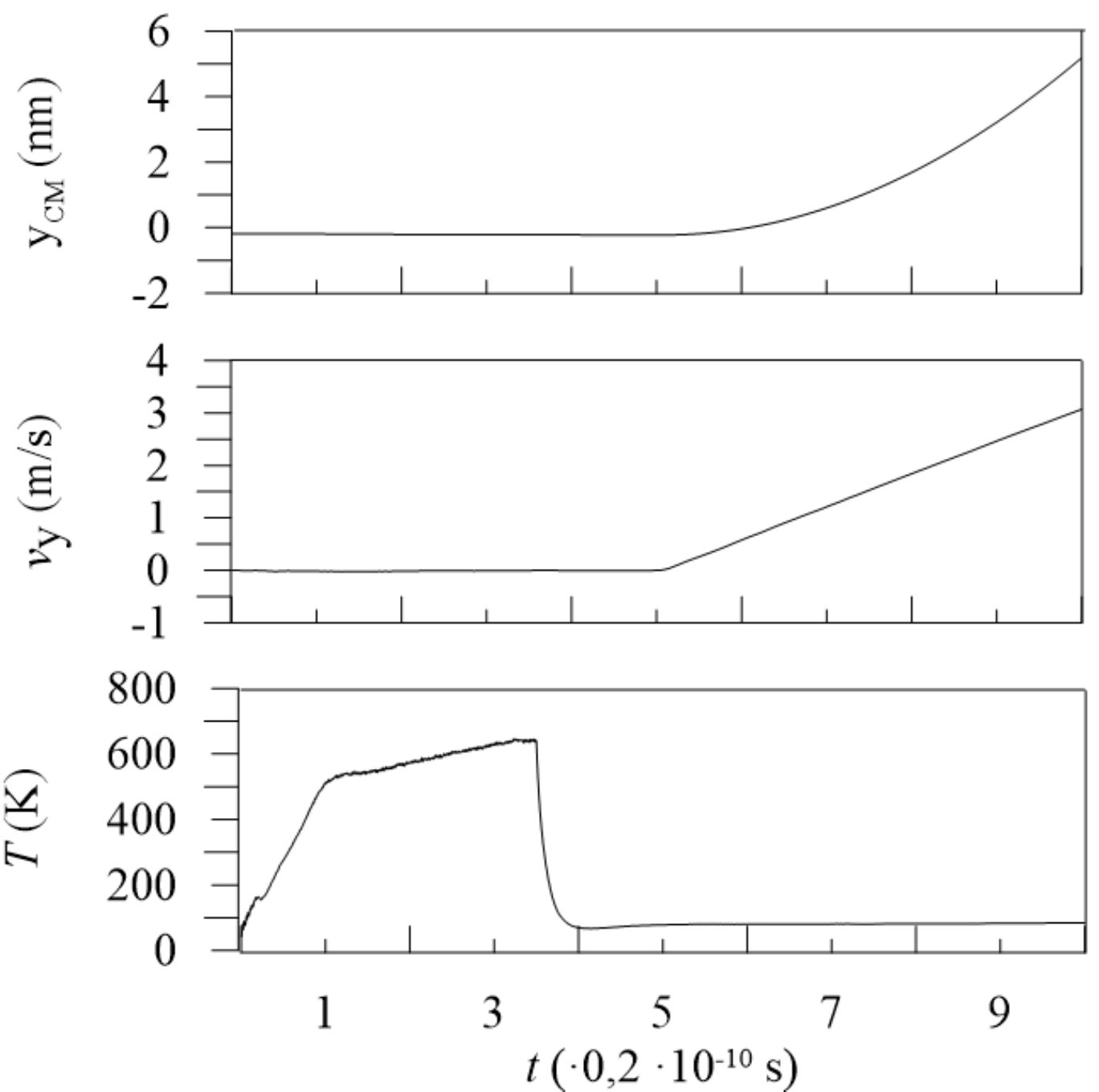}}
\caption{Time dependencies of quantities for nanoparticle at $\theta$ = $45^\circ$ and at a preset temperature of 80~K: $F_{sy}$ is the substrate force; $F_a$ is the applied force; $L_x$ and $L_y$ are the lateral sizes of nanoparticle in the $x$- and $y$-directions, respectively; $y_{CM}$ and $v_y$ are the $y$  components of the position and velocity of the nanoparticle center of mass, respectively; and $T$ is the system temperature.} \label{fig2}
\end{figure}

\begin{figure}[!t]
\centerline{\includegraphics[width=0.45\textwidth]{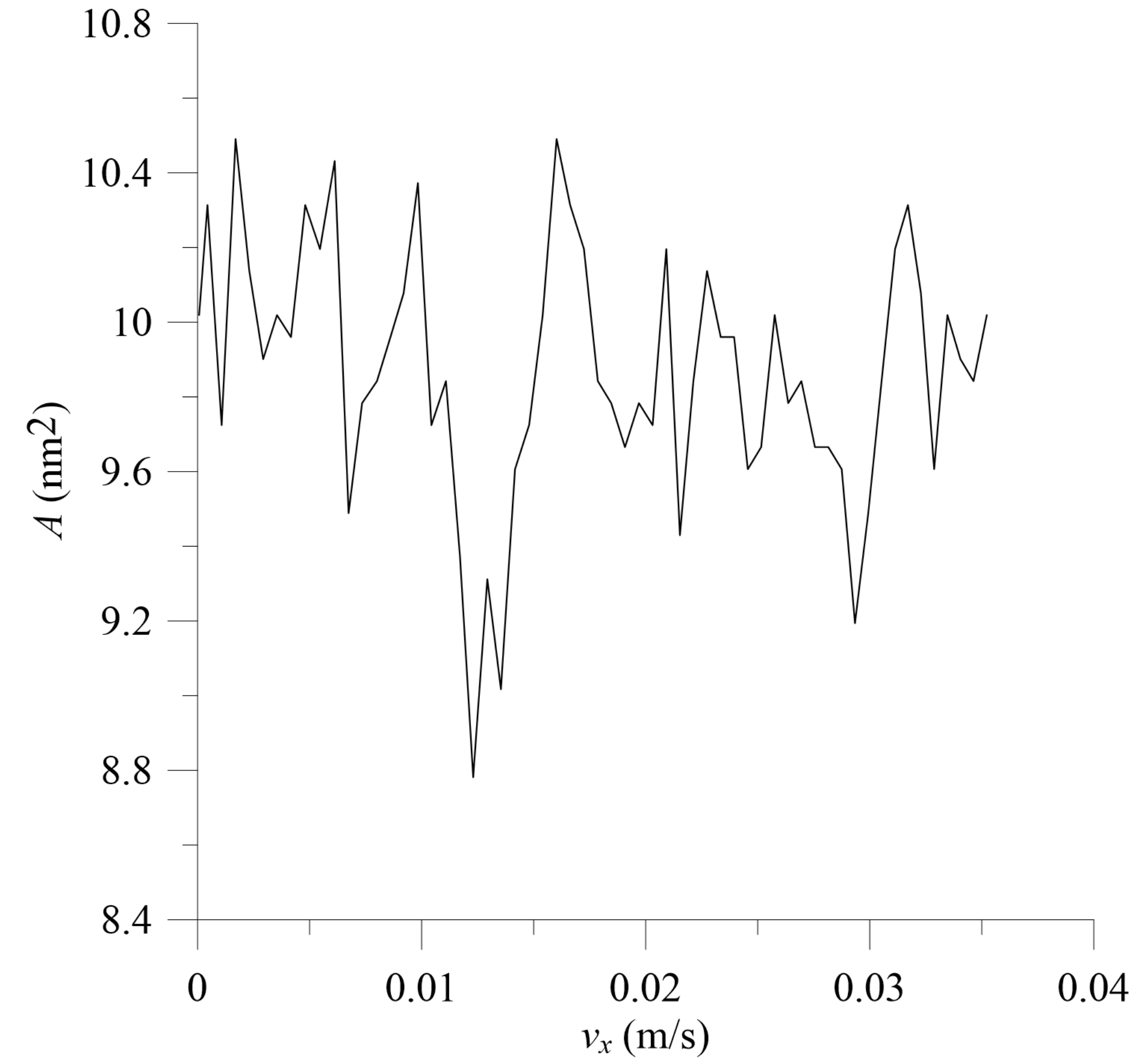}\includegraphics[width=0.45\textwidth]{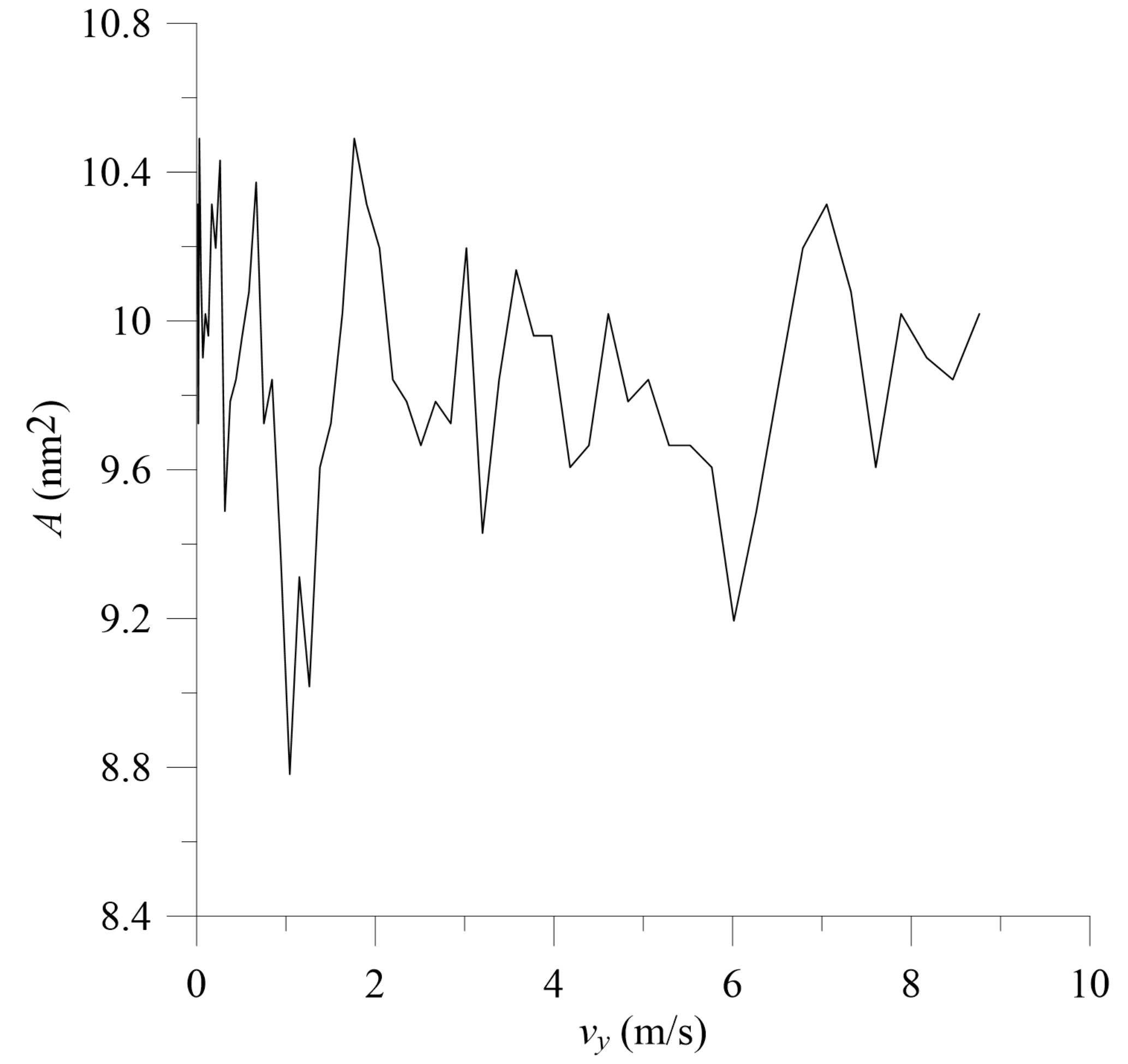}}
\caption{Dependencies of the contact area on the velocity along the $x$ (left) and $y$ (right) axes.} \label{fig3}
\end{figure}

Figure~\ref{fig2} is plotted for the stage of motion of nanoparticle at an angle $45^\circ$ and a temperature of 80 K. The maximum temperature of the system during simulation is 645 K, which is about three times lower than the melting point of palladium (1828 K). At the stage of formation and cooling of the nanoparticles, the speed of the nanoparticles is equal to zero. Speed increases throughout the movement. The contact area $A$ in figures~\ref{fig3} varies regardless of the velocity due to the deflection of the substrate under the nanoparticle. If we disable the application of force and nanoparticle does not move, the area is not changed due to the absence of substrate deformation.

\begin{figure}[!b]
\centerline{\includegraphics[width=0.45\textwidth]{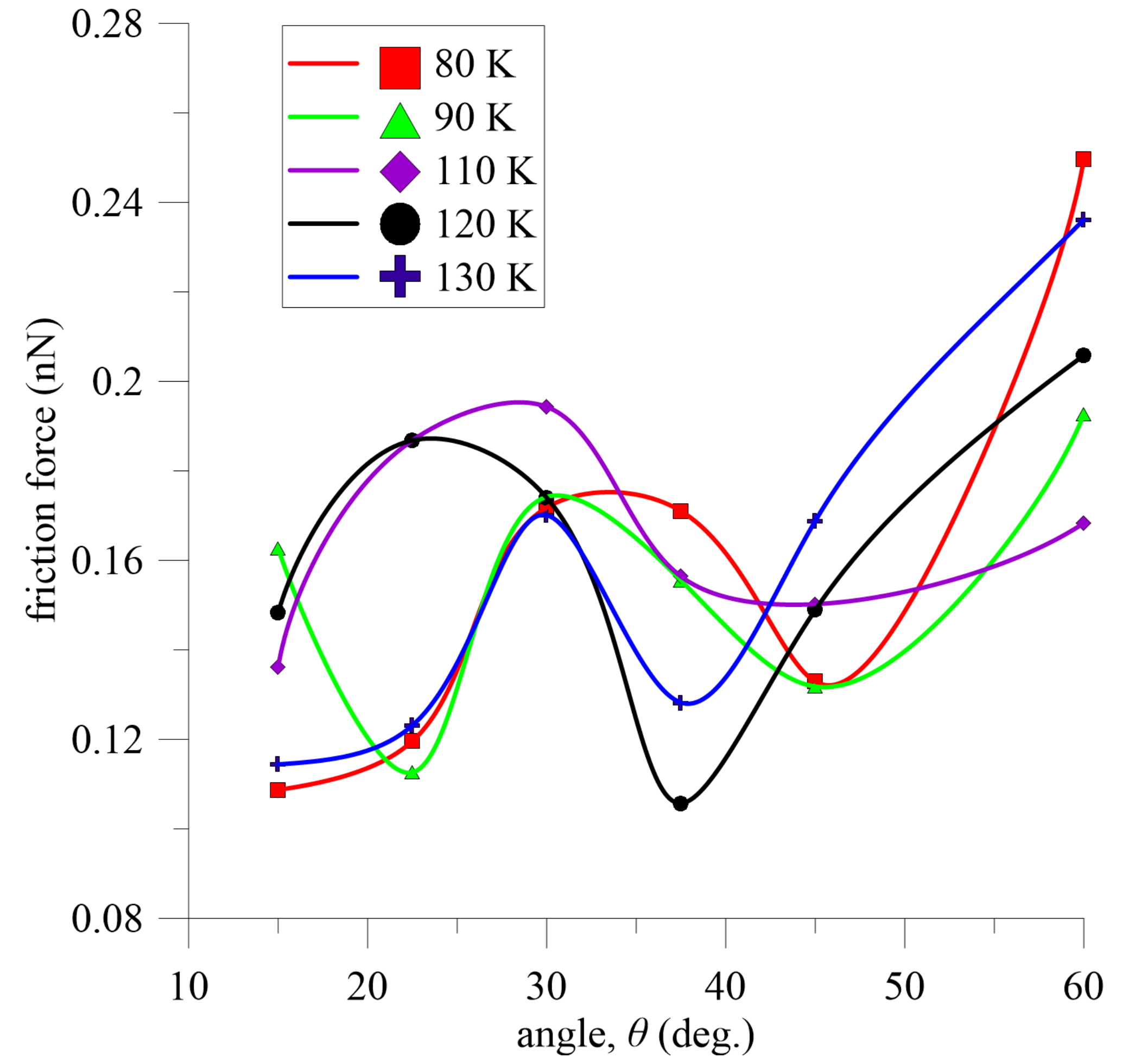}}
\caption{(Colour online) Anisotropy of the nanoparticles friction at temperatures of 80, 90, 110, 120 and 130 K.} \label{fig4}
\end{figure}

In comparison with our previous works \cite{TL_19_Khom,jnep_13}, we have chosen a much larger temperature step and range, in order to understand how the dependence of the friction force on the angle of the applied force varies due to the choice of the temperature. The markers in figure~\ref{fig4} correspond to the calculated average values of the friction force over the entire movement, and the solid lines are cubic splines. The figure shows that at angles of $22.5^\circ$ and $30^\circ$, the majority of nanoparticles at  different temperatures have approximately the same friction force. In most cases, the maximum value of the friction force is approximately two times greater than the minimum value. The dependencies of friction force at temperatures of 80 and 90 K jointly are characterized by minimum at an angle $45^\circ$ and starting from this value $F$ abruptly increases to a maximum magnitude with increasing angle. The curve describing anisotropy at 90~K has also a minimum at an angle of $22.5$ degrees. At the temperatures of 80 and 130 K, the minimal value of friction force is observed at an angle of $15^\circ$. When the system has a temperature of 120 K, a minimum of friction is observed at an angle of $37.5^\circ$. The friction force curves at 120, 130, and 110 K show a similar behavior, where the friction force increases first, and then decreases after the angles of $22.5$ and $30$ degrees and starts to grow again after the angle of $37.5^\circ$ and $45^\circ$. All dependencies in the figure show a similar nonmonotonous behavior.

\section{Conclusions}

We have come to a conclusion that the friction anisotropy is largely dependent on the surface and subsurface deformations that occur in selected directions during the slip process. Furthermore, microplastic deformations at the nanoparticle-graphene interface can be the cause of energy dissipation, but the main factor is apparently related to the surface tension that occurs under the contact pressure. It can be also concluded that the behavior of anisotropy also depends on the choice of the temperature. In other words, in structural superlubricating contacts, the rubbing forces are governed by compensations related to weakly compliant ideally crystalline incommensurate surfaces, which lead to the appearance of moire samples. Such regularities are connected with the changes of  tangential forces, in particular, during shear of nanoparticles over surfaces. In addition, friction is fixed basically by the boundary uncompensated regions of the soliton-type, which are located near the edges of the nanoparticles. The above-mentioned features define the fundamental and general properties of structural lubricity, namely, a weak dependence on the contact area and complex effects of the contact form and arrangement. 
\section*{Acknowledgements}

This investigation is supported by the Ministry of Education and Science of Ukraine within the framework of project ``Atomistic and statistical representation of formation and friction of nanodimensional systems'' (No.~0118U003584) and visitor grant of Forschungszentrum-J\"ulich, Germany. A.K. thanks Dr. Bo N.J.~Persson for hospitality during his stay in Forschungszentrum-J\"ulich.



%
%

\ukrainianpart

\title{Атомістичне моделювання фрикційної анізотропії наночастинок паладію на графені}
\author{О.В. Хоменко\refaddr{label1,label2}, М.В. Захаров\refaddr{label1}}
\addresses{
\addr{label1} Сумський державний університет, вул. Римського-Корсакова, 2, 40007 Суми, Україна
\addr{label2} Інститут Петера Грюнберга-1, Дослідницький центр Юліха, 52425, Юліх, Німеччина
}
%
%
%

\makeukrtitle

\begin{abstract}
\tolerance=3000%
Ця стаття є продовженням наших попередніх досліджень фрикційної анізотропії металевих наночастинок на поверхні графенової підкладки при інших температурних умовах. Сила тертя, що діє на наночастинки паладію на графеновому листі в різних бічних напрямках, вивчається за допомогою моделювання методом класичної  молекулярної динаміки. Анізотропія досліджується при високій швидкості ковзання наночастинок, що містять 10000 атомів, на поверхні графена. Ефект несумірності та ближній порядок контактних поверхонь наночастинок, а також деформація графена приводять до відсутності вираженої кутової залежності сили тертя.
\keywords атомно-силова мікроскопія, вуглецеві матеріали, тертя, графен, наночастинки, трибологія

\end{abstract}

\end{document}